# Holographic Methods as Local Probes of the Atomic Order in Solids

G. Faigel, M Tegze, G. Bortel, Z. Jurek, S. Marchesini*, M. Belakhovsky[o], A. Simionovici+

Research Institute for Solid State Physics and Optics, H-1525 Budapest, POB 49 Hungary,

*Lawrence Livermore National Laboratory 7000 East Ave., Livermore CA 94550, USA,

[o]CEA Grenoble/DRFMC-SP2M 17 Rue Des Martyrs, 38054 Grenoble France,

+ESRF BP 220, F-38043 Grenoble Cedex, France

**Abstract**

In the last fifteen years several techniques based on the holographic principle have been developed for the study of the 3D local order in solids. These methods use various particles: electrons, hard x-ray photons, gamma photons, or neutrons to image the atoms. Although the practical realisation of the various imaging experiments is very different, there is a common thread; the use of inside reference points for holographic imaging. In this paper we outline the basics of atomic resolution holography using inside reference points, especially concentrating to the hard x-ray case. Further, we outline the experimental requirements and what has been practically realized in the last decade. At last we give examples of applications and future perspectives.

**Introduction**

The knowledge of atomic and molecular structure is fundamental in physics, chemistry and biology. Beside the interest in basic research, there is also an increasing need from the high tech industry to control the structure of materials at the atomic level. This leads to the development of various methods, which can recover the microscopic structure of materials. Most of these techniques rely on the intensity measurement of the waves elastically scattered by the sample. However, the spatial arrangement of the scattering objects determines not only the intensity distribution but also the phase relation of waves traveling in different directions. The inversion of the intensity data to 3D real space atomic order is not straightforward without the phase. In many cases additional knowledge on the sample (such as composition, part of the atomic structure, closely related structures etc.) might replace the missing phase information and allow successful structure solution. However, there are problems, which hard to tackle with present techniques. Therefore methods giving direct phase information or data allowing the retrieval of phases are valuable. Holography using local reference point is one of these. In this paper we describe the basics of this method, show some examples for applications and outline future directions. After a general introduction we concentrate on those variants, which use hard x-rays for imaging.

**1. Holography using inside reference points**

Holographic imaging retains phase information by coherently mixing the scattered waves with a known reference wave. One of the most often used reference waves is the spherical wave ideally emitted by a point source. In practice the sources have finite sizes. This and the wavelength of the hologram forming waves determine the smallest feature size resolved in a given experimental arrangement. Therefore aiming atomic resolution both the source size and the wavelength of the radiation have to be in the angstrom range or smaller. This leads to the idea of using the atoms of the sample as sources of radiation [1,2]. In this case the structure is imaged relative to these atomic size reference points. Atoms can emit electrons, photons or incoherently scatter neutrons in the form of spherical waves. All these particles can be used for holographic imaging. The emitted radiation can reach the detector placed in the far filed, directly (without interacting with the atoms in the sample) or via elastic scattering on the surrounding atoms (Fig.1a). We can identify the former as the reference beam, the latter as the object beam. Interference between them results in angular dependent intensity, which is the hologram itself. In practice a macroscopic sample contains many source atoms producing independent holograms. These cannot be separated, which would result in a composite picture unusable for structure determination. This problem can be circumvented by using samples having source atoms with the same environment oriented in the same way [1].

Beside the above-described method, holograms can be taken in a different way. One can reverse the beam path using the optical reciprocity theorem; and the positions of the detector and source is exchanged [3]. In this case the atoms serve as local point detectrors and the hologram is formed by the incident plane waves and the waves scattered by the atoms surounding the detector atom (Fig.1b). In this case the hologam is measured by variing the incident beam direction compared to the sample and collecting the radiation emitted by the detector atom in the full solid angle. This method is often called inverse holography.

For clarity we summarize here the most important conditions to get a hologram

i. The environment of every source or detector atom has to be the same, and oriented in the same way

ii. The emitted radiation has to be monochromatic

iii. The waves emitted by different atoms have to be incoherent

iv. The size of the sample has to be much smaller than the sample-detector distance.

In this case separate but identical holograms are simply added.

**2. Comparison of holographic techniques**

Since the scattering processes for electrons, photons and neutrons differ significantly, the image formation is also different and the inversion of the measured data to real space structure needs special treatment in all these cases. Below we give a brief comparison of holographic methods using various hologram-forming waves.

Electrons scatter strongly on atoms. This results in large holographic oscillations. Therefore it is relatively easy to measure electron holograms. However, the strong scattering leads to large object-object interference and also to multiple scattering. Further the scattering amplitudes and phases are strongly angular dependent. All of these effects distort the holographic information. Therefore, many efforts were concentrated to compensate for these [4-6]. From the experimental point of view electron holography is similar to electron diffraction. It needs ultra vacuum, clean surface, conducting sample. It is used to get structural information on crystal surfaces. In the last ten years there have been many publications on various applications [7-10].

The interaction of hard x-ray photons with matter is much weaker than that of the electrons. The elastic scattering on the atomic electrons is more isotropic and the phase shift can be neglected in most of the cases. Therefore the holographic image is much closer to the ideal than in the case of electrons. No corrections are necessary to correct the effects of multiple scattering and phase shift. The weak interaction allows the photons to penetrate deeply into the sample. Therefore bulk materials can be studied. However, the small elastic scattering amplitude leads to small holographic oscillations on a high continuous background. Therefore images with very good statistics have to be taken, and special care is needed to avoid any systematic errors, which would mask or distort the holographic signal. These experimental problems limit hard x-ray holographic studies to large flat single crystals [11]. Though at synchrotron sources the high incident flux at undulator beamlines allows the measurement of thin layers, and smaller samples. At the next section we describe some of the experimental results in details.

A special type of hard x-ray holography is the gamma ray holography. In this case the energy of the incident beam coincides with a low-lying nuclear level, which leads to resonant scattering of photons on the nuclei. Since the scattering factor of this process is usually larger than scattering of photons by atomic electrons, the holographic oscillations are also larger. So one would expect easier measurement. However, the very narrow energy bandwidth of the resonant scattering makes these measurements difficult. A further complication is the need for special isotopes, which have the proper nuclear levels. In spite of these difficulties there are efforts to carry out this type of experiments [12]. The reason is that beside the geometrical arrangement of atoms they could also give information on the magnetic structure, owing to the hyperfine field dependence of the nuclear scattering factors [13].

The last hologram forming waves we discuss are neutrons. Generally they interact weakly with matter. Therefore their penetration is deep. This combined with the large (external) source size leads to the need of large bulk samples. Since the elastic scattering amplitude is small, the magnitude of holographic oscillations is also small. Experiments are also hindered by the low intensity of existing neutron sources. Therefore the measurement of neutron holograms is difficult. There have been only two experiments, both of them with unconvincingly poor statistics [14,15]. On the other hand the systematic distorting effects, such as angular dependence of elastic scattering factor, multiple scattering, phase shift are the smallest in the case of neutrons. A further advantage is the easier imaging of some of the light elements.

## 2. Hard x-ray holography with fluorescent radiation

In this section we present the basics of hard x-ray holography and a few examples of applications. The easiest and most often used way to form a hologram by hard x-ray photons is the use of K fluorescence of medium Z elements (Mn to Mo). This process can be initiated by an external x-ray source having energy larger them the K absorption edge of the fluorescing atom. In normal mode the fluorescent photon propagates in the sample as a spherical wave. It can reach the detecting surface (a hemisphere) directly or via scattering on the neighbors of the emitter. These two waves are coherent so they interfere. The spatial variation of the intensity on the detector surface is the hologram. Note that we neglect multiple scattering and the object-object term. This is justified by the small cross section of elastic scattering of hard x-ray photons on the atomic electrons. In the inverse mode we also measure the fluorescent radiation, however not its angular variation but its integrated value over the full solid angle. In this case the incident beam forms the hologram. This plane wave can reach the "detector atom" (the fluorescing element) directly or via scattering on its neighbors. These two waves interfere and the intensity of the interference field varies as we change the direction of the incident radiation relative to the sample. The total number of fluorescent photons expected to be proportional to the field strength, reflecting the holographic oscillations. A typical setup working both in normal and inverse mode is depicted in Figure 2. The two vertical rotations θ and θ'are used in normal and inverse mode respectively. The horizontal ϕ rotation does full turns in both modes. This way a scanning of the detector (or the source in the inverse mode) on a hemisphere is mimicked.

At last we would like to show two applications: The first one is the imaging of the atomic decoration in a PdAlMn quasicrystal [16]. As it is well known, quasicrystals are non-periodic in 3D, but in 6D they can be described as a periodic lattice [17]. However, projecting this lattice to 3D does not give the atomic decoration. This has to be modelled starting from chemical considerations. Local methods, such as electron microscopy or AFM could give a picture of the atomic order. However, they probe only a small area on the surface of the sample. To get a picture of the atomic order in the bulk, x-ray methods have to be used. Though traditional crystallographic measurements show strong peaks in well-defined directions, the atomic positions cannot be derived. Using holography we could image selected shells about the Mn atoms. The hologram of the environments of Mn atoms is shown in Fig. 3a and the reconstructed image in Fig. 3b. Comparing the result to model structures, one could validate theoretical predictions [16]. The second example is a study of the environment of Zn dopant in a GaAs matrix [18]. The holograms at 9.7 keV and at 10 keV are shown in Fig. 4a and 4b, respectively. Three reconstructed planes are depicted in Fig. 5a-c. The first one contains the Zn central atoms (a), the next shows the plane 1.41 Å above (b) and the last 1.41 Å below (c). Quantitative analysis shows that Zn atoms substitute Ga atoms selectively.

At last we would like to mention one more measurement, the imaging of the 3D order of atoms in a NiO crystal; Fig. 6 [19]. Though the NiO has a well know structure, the holographic studies of this sample attracted attention. The reason for this is the succesful reconstruction of the oxygen atomic positions. Since the amplitude of the holographic oscillations is proportional to the scattering factor of the scatterer, the imaging of light atoms is difficult. In all of the previous studies only the position of heavy elements could be reconstructed. This experiment gives hope to extend the class of samples good for holographic imaging to a much wider range, even to biological materials.

## 3. Conclusions

Introducing a new method, the first question one asks is: What can this method give us, which other techniques cannot?

In the case of holography the answer is as follows: hard x-ray holography is a local method such as EXAFS, but unlike EXAFS, which yields the pair correlation function, holography gives the direct 3D order of atoms about a selected element in the sample. It does not suffer from the phase problem as diffraction. Further, in principle it is capable of imaging non-periodic objects. This might be especially useful in experiments at the future fourth generation x-ray sources. These features make holography an attractive method for structural studies. Of course, there are many technical problems, which have to be solved before wide application. In this article we have given a brief comparison of inside reference points based holographic techniques using various hologram-forming waves. We have shown a few applications, which illustrated the potential power of the method.

**Acknowledgment**

This work was supported by OTKA T043237, and T034284.

**Figures**

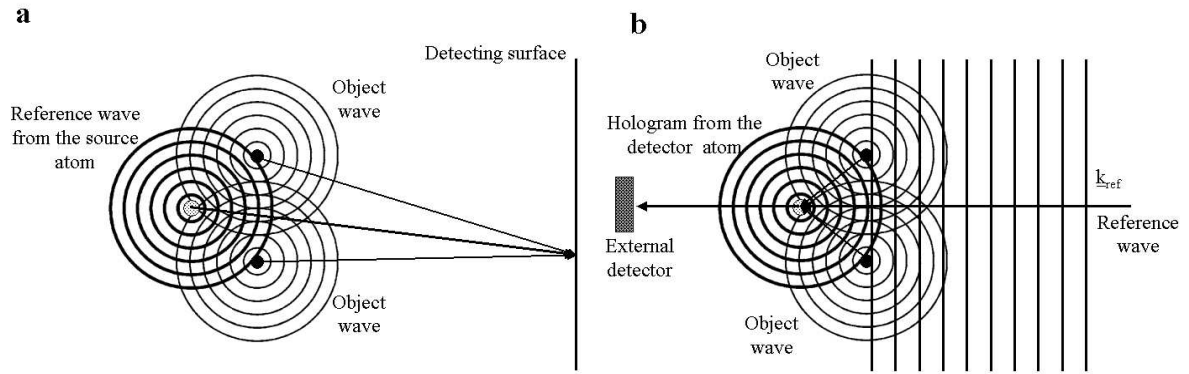

1. Two basic arrangements of holography using inside reference points: (a) inside source holography, (b) inside detector holography.

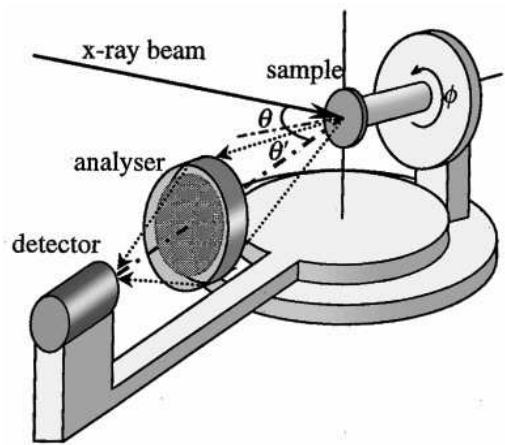

2. The sketch of the experimental setup used in x-ray fluorescent holography.

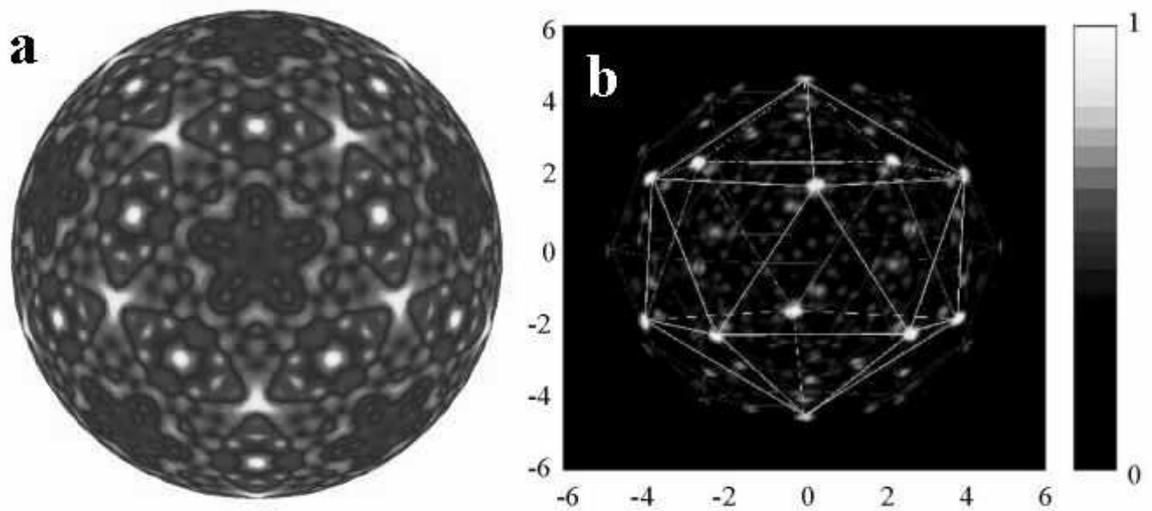

3. Holographic imaging of an $Al_{70.4}Pd_{21}Mn_{8.6}$ quasi crystal, (a) hologram of the environment of Mn atoms, (b) reconstructed image of selected coordination shells about the Mn atoms.

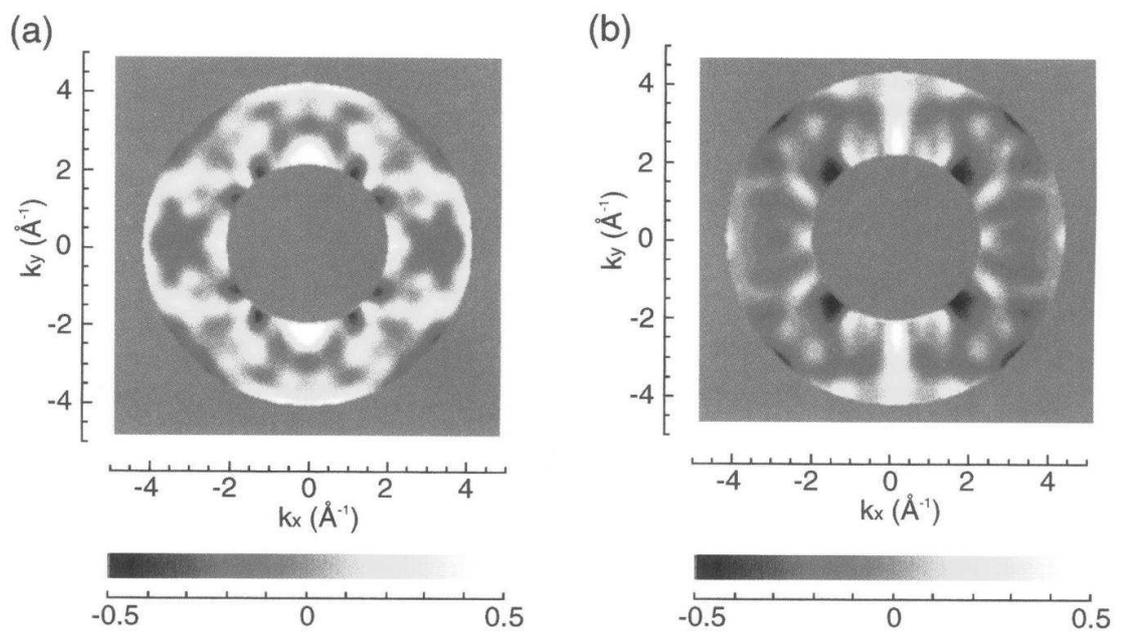

4. Holograms of the atomic environment of the Zn atoms in a GaAs:Zn sample at 9.7 (a) and 10 keV (b).

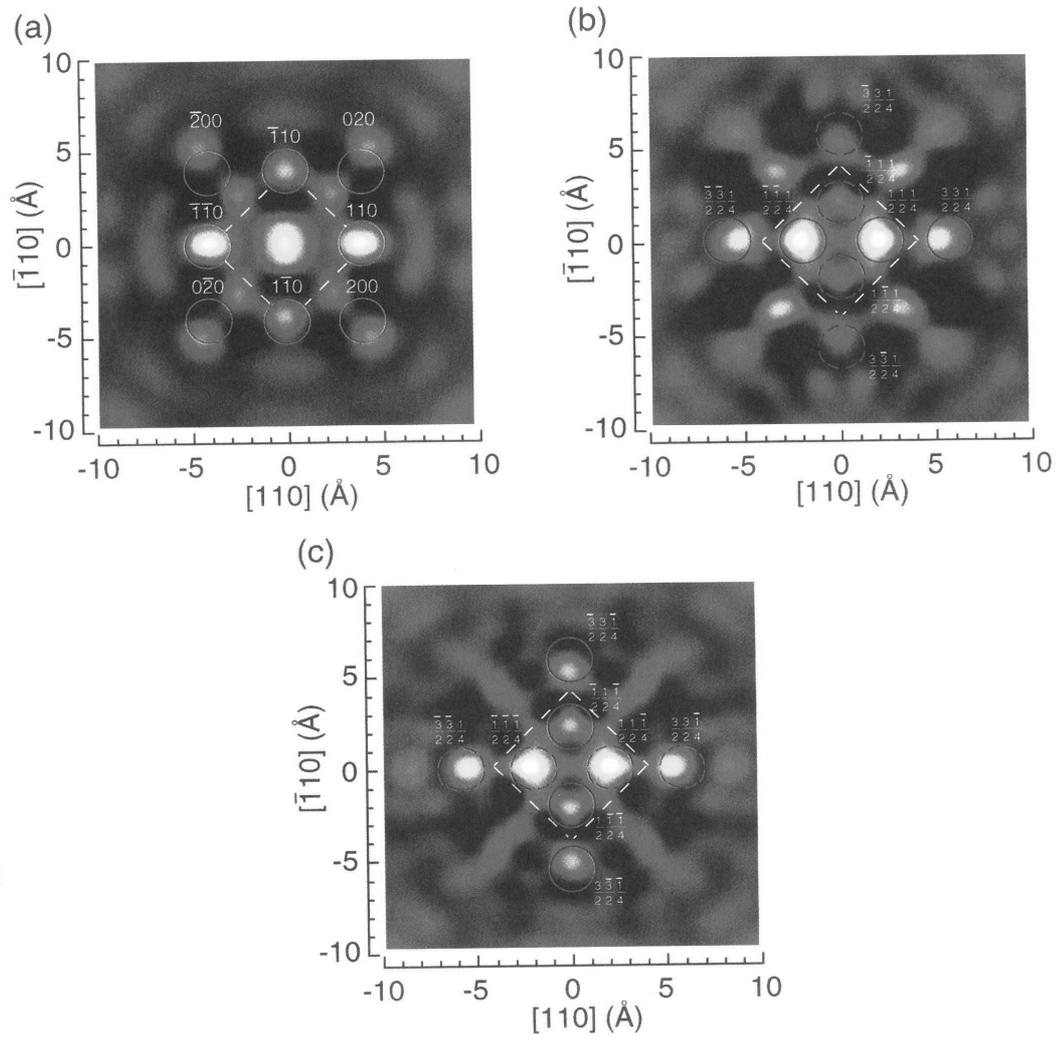

5. Three reconstructed planes of the GaAs sample, the plan containing the source atom (a), one below by 1.4 Å (b), and one above by 1.4 Å (c).

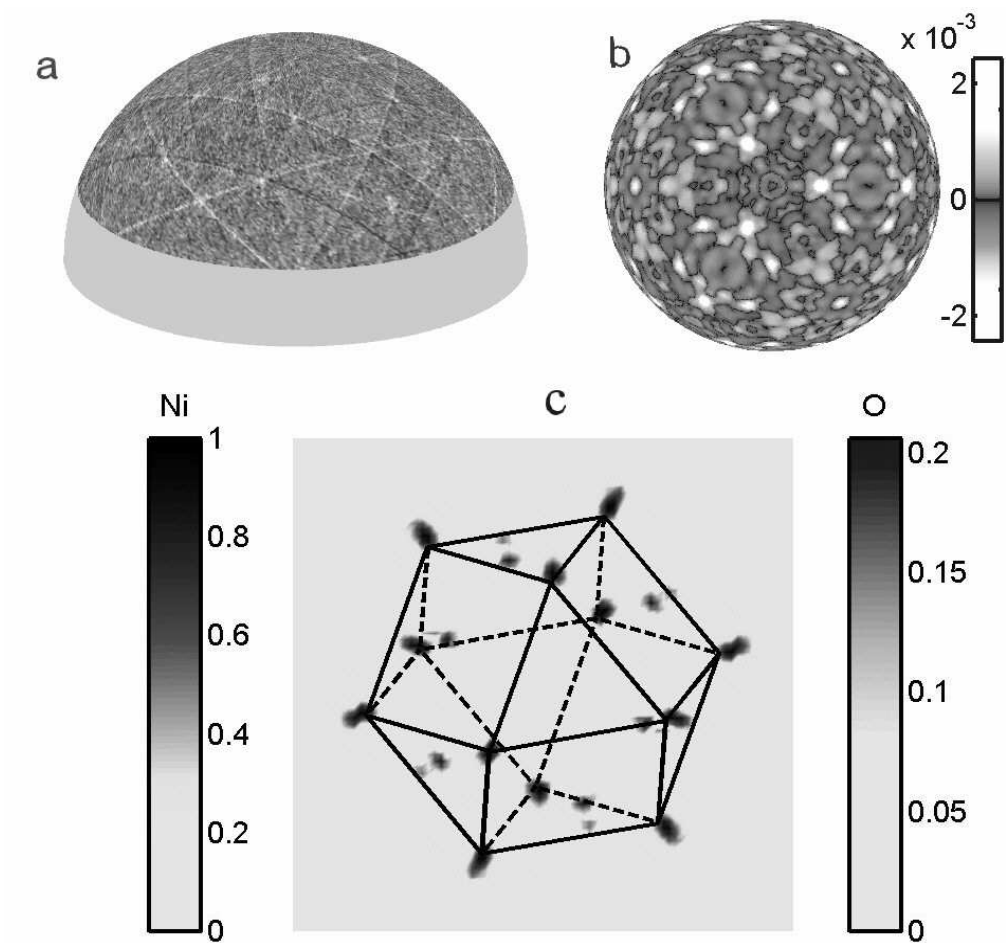

6. Holographic imaging of the Nickel and Oxygen atoms in a NiO sample: background corrected data (a), full hologram keeping only the contribution of the first few atomic shells about the Ni atoms (b), reconstructed real space image (c).